\documentclass[letterpaper, 10 pt, conference]{ieeeconf}  

\IEEEoverridecommandlockouts                              

\usepackage{amsmath}
\usepackage{amsfonts}
\usepackage{graphicx}
\usepackage{colortbl}
\usepackage{cite}
\usepackage{algorithm}
\usepackage{algpseudocode}

\newtheorem{theorem}{Theorem}[section]

\newtheorem{remark}[theorem]{Remark}

\makeatletter
\let\NAT@parse\undefined
\makeatother
\usepackage{hyperref}       

\title{\LARGE \bf
Dual State-space Fidelity Blade (D-STAB): A Novel Stealthy Cyber-physical Attack Paradigm
}

\author{Jiajun Shen$^{1}$, Hao Tu$^{1}$, Fengjun Li$^{2}$, Morteza Hashemi$^{2}$, Di Wu$^{3}$, and Huazhen Fang$^{1}$
\thanks{$^{1}$Jiajun Shen, Hao Tu and Huazhen Fang are with the Department of Mechanical Engineering, The University of Kansas, Lawrence, KS 44906, USA
        {\tt\small sjjvic@gmail.com, tuhao, fang@ku.edu}}%
\thanks{$^{2}$Fengjun Li and Morteza Hashemi are with the Department of Electrical Engineering and Computer Science, The University of Kansas, Lawrence, KS 44906, USA
        {\tt\small fli, mhashemi@ku.edu}}%
\thanks{$^{3}$Di Wu is with Pacific Northwest National Laboratory (PNNL), 902 Battelle Blvd, Richland, WA 99354, USA
        {\tt\small Di.Wu@pnnl.gov}}%
}

\begin{document}

\maketitle
\thispagestyle{empty}
\pagestyle{empty}

\begin{abstract}
This paper presents a novel cyber-physical attack paradigm, termed the Dual State-Space Fidelity Blade (D-STAB), which targets the firmware of core cyber-physical components as a new class of attack surfaces. The D-STAB attack exploits the information asymmetry caused by the fidelity gap between high-fidelity and low-fidelity physical models in cyber-physical systems. By designing precise adversarial constraints based on high-fidelity state-space information, the attack induces deviations in high-fidelity states that remain undetected by defenders relying on low-fidelity observations. The effectiveness of D-STAB is demonstrated through a case study in cyber-physical battery systems, specifically in an optimal charging task governed by a Battery Management System (BMS).
\end{abstract}

\section{INTRODUCTION}

\subsection{Cyber-physical Attacks: Background and Overview}

The rapid advancement of technology has accelerated the evolution of cyber-physical systems (CPS), seamlessly integrating computational and physical components across industries such as transportation, healthcare, smart grids, and industrial automation \cite{humayed2017cyber}. While this integration significantly enhances efficiency, flexibility, and automation capabilities, it simultaneously broadens the attack surface, rendering CPS vulnerable to cyber-physical threats.

Common cyber-physical attack strategies include Denial of Service (DoS) \cite{li2020active, shen2020cross}, False Data Injection (FDI) \cite{weng2023secure}, and replay attacks \cite{liu2021active, sanchez2019detection}. These methods disrupt CPS operations by targeting data exchanges between physical devices and controllers. For instance, DoS attacks saturate communication channels to disrupt command execution; FDI attacks inject manipulated data to misguide control decisions; replay attacks retransmit valid but outdated data to confuse system states and control actions.

\subsection{Shortcomings of Current Attack Paradigms}

Despite extensive research, existing cyber-physical attacks face practical limitations. Firstly, traditional attack surfaces, such as sensors, actuators, and communication channels, have become increasingly inaccessible due to advanced semiconductor integration (e.g., System-on-Chip and MEMS sensors). Modern systems, like integrated Battery Management Systems (BMS), embed communication channels within multi-layered printed circuit boards, making physical and electronic access challenging without sophisticated methods like electromagnetic interference (EMI), which themselves are mitigated by advanced shielding techniques \cite{zhang2021survey, lucia2022supervisor, dukosi, dayanikli2020electromagnetic}.

Secondly, existing attacks are typically resource-intensive and economically inefficient. DoS attacks require sustained high-volume network traffic, while FDI attacks necessitate real-time data manipulation and spoofing, both demanding substantial computational resources. Replay attacks are limited by the constraints of pre-captured data, significantly reducing their flexibility and effectiveness in dynamic scenarios \cite{li2020active, weng2023secure, liu2021active}.

Thirdly, these attacks often lack sophisticated domain-specific intelligence, making them predictable and easily detectable. They typically do not consider system dynamics or real-time context, reducing their potential impact and simplifying detection by modern anomaly monitoring systems. Finally, traditional attacks often fail stealth criteria, easily identifiable through abnormal data patterns, traffic anomalies, or repeated data transmissions detected by advanced security algorithms prevalent in current infrastructures \cite{sanchez2019detection, mavikumbure2024cy}.

\subsection{D-STAB: A Stealthy Attack Paradigm via Fidelity-gap}

To address these limitations, this paper introduces a novel cyber-physical attack paradigm: the Dual State-Space Fidelity Blade (D-STAB). D-STAB uniquely exploits the fidelity gap—discrepancies between high-fidelity and low-fidelity system models—leveraging firmware parameters stored within controllers (e.g., ROM, EEPROM, flash memory). By precisely manipulating these parameters using advanced domain knowledge, D-STAB achieves stealthy and efficient disruption.

Distinct advantages of the D-STAB paradigm include:

\begin{itemize}
    \item {\bf Ease of Implementation}: D-STAB can be executed through both physical and remote access methods. Physical access is enabled using common tools such as JTAG, J-Link, or ISP interfaces, which allow direct read/write operations on the firmware, facilitating parameter manipulation. Remotely, network interfaces like CAN and UART can be exploited via buffer overflows or malware, enabling attackers to alter system parameters or inject malicious code.

    \item {\bf Resource Efficiency}: As a firmware-based attack, D-STAB requires only minimal adjustments to a few key parameters to significantly disrupt system behavior. Unlike traditional attacks, such as FDI or DoS, which require sustained interference, D-STAB operates with a one-time access strategy. A single, well-placed modification to critical parameters can cause the system to deviate from normal operation, achieving maximal disruption with minimal resource consumption.

    \item {\bf Context Awareness}: D-STAB is inherently context-aware and target-specific, leveraging detailed domain knowledge of the system. With access to the high-fidelity state-space model, attackers can identify and manipulate parameters that control critical processes. By carefully adjusting these parameters, the attack disrupts the system's control mechanisms in a precise and efficient manner, minimizing the need for prolonged intervention.

    \item {\bf Stealthiness}: D-STAB's stealthiness arises from its subtle firmware modifications, making it extremely difficult to detect. Unlike traditional attacks, such as DoS, FDI, or replay attacks, which are noticeable due to their disruptive nature or observable anomalies, firmware-based attacks exploit the state-space fidelity gap. The system, typically relying on low-fidelity models, lacks the necessary information to detect the attacker's high-fidelity modifications. This information asymmetry enables the attacker to introduce malicious changes that are indistinguishable from normal operations, rendering standard detection mechanisms ineffective.
\end{itemize}

\subsection{Fidelity-gap Exploitation: The Core Principle of D-STAB}
\label{subsection: Pillar of D-STAB}

\begin{figure}[htbp]
\centering
\includegraphics[width=0.48\textwidth]{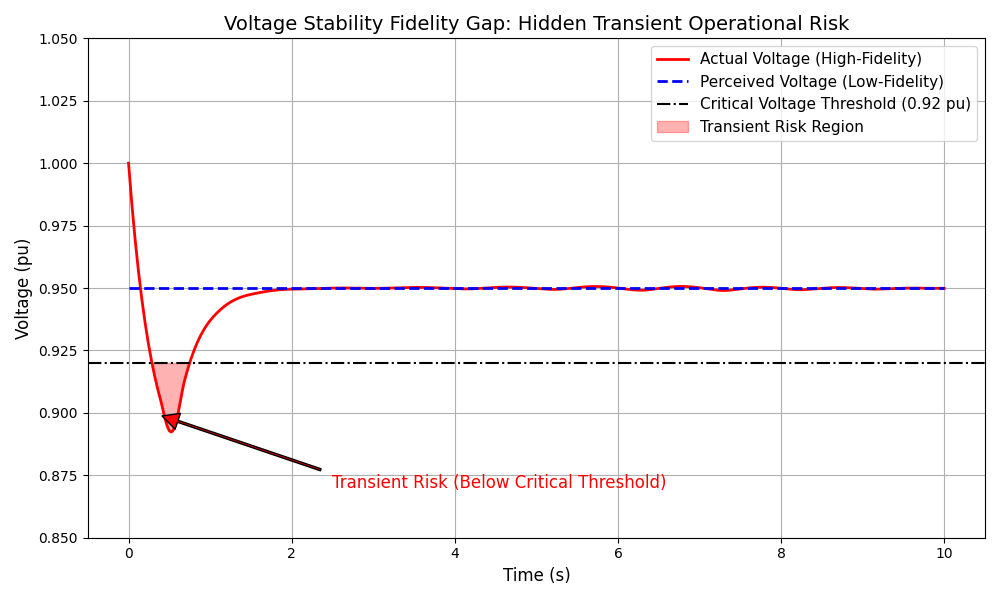}
\caption{Voltage Stability Fidelity Gap: Hidden Transient Risk Induced by Model Granularity Differences}
\label{fig: Fidelity Gap Voltage Stability Example}
\end{figure}

Voltage stability assessments in power system operations frequently rely on simplified, computationally efficient low-fidelity models. These models approximate system responses to disturbances as instantaneous steady-state conditions, neglecting critical transient behaviors. To explicitly illustrate the risks posed by this fidelity gap, we consider a practical scenario involving transient voltage dynamics after a short-term load disturbance, as shown in Fig. \ref{fig: Fidelity Gap Voltage Stability Example}.

In this scenario, the low-fidelity model (blue dashed line) predicts an immediate steady-state equilibrium at 0.95 pu, considered safe within the standard operational limits (±5\% range: 0.95–1.05 pu). Conversely, the high-fidelity model, which incorporates detailed transient dynamics through inertia and damping, reveals a significant transient voltage drop below the IEEE-defined critical threshold of 0.92 pu, marked by the shaded red region.

This example highlights the hidden risks from fidelity gaps—brief but severe transient voltage dips can trigger protective relays, unintended load shedding, and potentially cascading failures. Thus, relying solely on simplified steady-state models for operational decisions can inadvertently expose power grids to severe transient risks, emphasizing the importance of incorporating high-fidelity dynamic analyses in operational assessments.

\subsection{Main Contributions}

The main contributions of this work are two-fold. First, we propose the D-STAB, a novel and stealthy cyber-physical attack paradigm. This paradigm leverages a coupled framework, consisting of a high-fidelity adversarial constrained optimal control problem (OCP) and a low-fidelity inverse constrained OCP. This approach allows for precise manipulation of system behavior while remaining undetected by the defender relying on low-fidelity models. Second, we demonstrate the effectiveness of the proposed attack paradigm through its application in cyber-physical battery systems. The case study showcases how D-STAB can exploit the fidelity gap in real-world systems, manipulating battery charging processes while evading detection.

\section{D-STAB Attack Framework}



To exploit the state-space fidelity gap and overcome the limitations of existing attack paradigms, we propose a novel stealthy cyber-physical attack paradigm, the D-STAB. This paradigm is structured around a comprehensive framework that couples a high-fidelity adversarial constrained optimal control problem (OCP) with a low-fidelity inverse constrained OCP. Both OCPs are represented using the Model Predictive Control (MPC) formulation to ensure generality and applicability in real-time scenarios. This representation allows for iterative optimization and real-time decision-making, which accommodates the dynamic nature of practical cyber-physical systems.

\subsection{High-fidelity Advesarial MPC}

To execute the D-STAB attack, the attacker first defines a specific adversarial objective and designs corresponding constraints in the high-fidelity state-space model, while ensuring that the constraints perceived by the defender reflect normal system operations. These high-fidelity adversarial constraints are crafted based on domain-specific knowledge, leveraging the attacker's information advantage over the defender.

The attacker then solves the high-fidelity adversarial MPC problem to generate a sequence of adversarial control inputs. The high-fidelity adversarial MPC problem is formulated as follows:
\begin{equation}
    \label{eq: High fidelity adversarial MPC}
    \begin{aligned}
        & \min_{u_{0:N-1|t}} J(x_{H, t}; \theta_H) := \sum_{k=0}^{N-1} g_H(x_{H, k|t}, u_{k|t}; \theta_H) \\
        & + g_{H, N}(x_{H, N|t}; \theta_H) \\
        & \text{subject to:} \ x_{H, k+1|t} = f_{H, k|t}(x_{H, k|t}, u_{k|t}), \ x_{H, 0|t} = x_{H, t}, \\
        & w_A(x_{H, k|t}) \geq 0, \\
        & w_{H,k|t}(x_{H, k|t}, u_{k|t}) \leq 0  \ \text{for} \ k=0,1,\cdots,N-1 \\
        & w_{H,N|t}(x_{H, N|t}) \leq 0,
    \end{aligned}
\end{equation} where $x_{H, k|t} \in \mathbb{R}^{n_H}$ represents the high-fidelity system state at time step $t$, and prediction step $k$, and $u_{k|t} \in \mathbb{R}^{m}$ is the control input. The function $f_{H, k|t}: \mathbb{R}^{n_H} \times \mathbb{R}^{m_H} \rightarrow \mathbb{R}^{n_H}$ denotes the high-fidelity system dynamics, while $g_{H}: \mathbb{R}^{n_H} \times \mathbb{R}^{m_H} \rightarrow \mathbb{R}$ and $g_{H, N}: \mathbb{R}^{n_H} \rightarrow \mathbb{R}$ are the instantaneous and terminal cost functions parameterized by $\theta_H \in \mathbb{R}^{r_H}$. The constraint $w_A: \mathbb{R}^{n_H} \rightarrow \mathbb{R}$ represents the attacker-designed conditions that reflect specific adversarial goals, and $w_{H, k|t}: \mathbb{R}^{n_H} \times \mathbb{R}^{m_H} \rightarrow \mathbb{R}$ and $w_{H, N|t}: \mathbb{R}^{n_H} \rightarrow \mathbb{R}$ are constraints ensuring stealth by maintaining normal operations in the low-fidelity state-space representation.

\begin{remark}
    The low-fidelity system state $x_{k|t} \in \mathbb{R}^{n}$ can be retrieved from the high-fidelity state $x_{H,x|t}$ using a domain-specific mapping function $\psi: \mathbb{R}^{n_H} \rightarrow \mathbb{R}^{n}$, such that $x_{k|t} = \psi(x_{H,k|t})$. The D-STAB attacker, equipped with this knowledge, exploits this relationship to achieve specific malicious objectives by designing high-fidelity constraints $w_A(x_{H,k|t}) \geq 0$ while remaining undetectable to the defender. Due to the information asymmetry, where the low-fidelity decision-maker (defender) cannot access high-fidelity state information, the attacker's activities in the high-fidelity dynamics remain invisible, making these attacks nearly impossible to detect.
\end{remark}

\subsection{Low-fidelity Advesarial MPC}

Given the solution to the high-fidelity adversarial OCP, denoted as $u_{0:T}^{\text{adv}}$, the next challenge is determining how to adjust the parameters of the low-fidelity constrained OCP such that its optimal solution aligns with the adversarial control sequence. To address this, we formulate a bi-level optimization framework where the upper-level problem minimizes the discrepancy between the low-fidelity OCP solution and the high-fidelity adversarial control sequence. This upper-level problem is constrained by the lower-level problem, which consists of the low-fidelity constrained OCP.

The low-fidelity inverse MPC is defined as:
\begin{equation}
    \label{eq: Low-fidelity inverse constrained OCP}
    \begin{aligned}  
        & \min_{\theta} 
        L(u_{0:T}^*(\theta), u_{0:T}^{\text{adv}}) := \sum_{t=0}^T l(u_t^*(\theta),  u_t^{\text{adv}})\\
        & \text{subject to:} \ u_t^*(\theta) = u_{0|t}^*(\theta) 
    \end{aligned}
\end{equation}
where $u_{0|t}^*(\theta) \in \mathbb{R}^{m}$ is the first control input from the low-fidelity MPC solution, and $\theta \in \mathbb{R}^{r}$ are the parameters of the low-fidelity MPC problem \eqref{eq: Low-fidelity constrained OCP} defined below. The function $l: \mathbb{R}^{m} \rightarrow \mathbb{R}$ is the loss function measuring the discrepancy between $u_{0|t}^*(\theta)$ and $u_{0:T}^{\text{adv}}$. 

The low-fidelity MPC problem is defined as
\begin{equation}
    \label{eq: Low-fidelity constrained OCP}
    \begin{aligned}
        & \min_{u_{0:N-1|t}} J(x_t; \theta) := \sum_{k=0}^{N-1} g(x_{k|t}, u_{k|t}; \theta) + g_{N}(x_{N|t}; \theta) \\
        & \text{subject to:} \ x_{k+1|t} = f_{k|t}(x_{k|t}, u_{k|t}), \ x_{0|0} = x_{\text{init}}, \ x_{0|t} = x_t, \\
        & w_{k|t}(x_{k|t}, u_{k|t}) \leq 0, \ \text{for} \ k=0,1,\cdots,N-1 \\
        & w_{N|t}(x_{N|t}) \leq 0,
    \end{aligned}
\end{equation} where $x_{k|t} \in \mathbb{R}^{n}$ and $u_{k|t} \in \mathbb{R}^{m}$ are the low-fidelity system state and control input, respectively. The function $f_{k|t}: \mathbb{R}^{n} \times \mathbb{R}^{m} \rightarrow \mathbb{R}^{n}$ denotes the low-fidelity system dynamics, and $g: \mathbb{R}^{n} \times \mathbb{R}^{m} \rightarrow \mathbb{R}$ and $g_{N}: \mathbb{R}^{n} \rightarrow \mathbb{R}$ are the instantaneous and terminal cost functions respectively, parameterized by $\theta \in \mathbb{R}^{r}$. Constraints $w_{k|t}: \mathbb{R}^{n} \times \mathbb{R}^{m} \rightarrow \mathbb{R}$ and $w_{N|t}: \mathbb{R}^{n} \rightarrow \mathbb{R}$ ensure the normal operation in the low-fidelity state-space.


\subsection{Gradient-based Optimization Approach}

Having defined both the high-fidelity and low-fidelity MPC formulations, the next step is to construct the inverse MPC as described in \eqref{eq: Low-fidelity inverse constrained OCP}. The goal is to derive the ``optimal'' malicious parameters to inject into the firmware, such that the compromised low-fidelity MPC yields a solution as close as possible to the adversarial control sequence generated by the high-fidelity adversarial MPC. To achieve this, we define the discrepancy loss function as $l(u_t^*(\theta),  u_t^{\text{adv}}) = \| u_t^*(\theta) -u_t^{\text{adv}} \|^2$.

Given this bi-level optimization problem, there are several methods to solve it, as discussed in \cite{liu2021investigating}. One way is to adopt a computationally efficient, gradient-based approach:
\begin{equation*}
    \theta_{i+1} = \theta_i - \alpha \frac{\text{d} L}{\text{d} \theta} \Big|_{\theta_i},
\end{equation*}
where $i$ denotes the iteration index, $\alpha$ is the learning rate, and $\frac{\text{d} L}{\text{d} \theta} \Big|_{\theta_i}$ is the gradient value of the loss w.r.t. $\theta$ evaluated at $\theta_i$. The gradient term $\frac{\text{d} L}{\text{d} \theta} \Big|_{\theta_i}$ can be computed based on the Pontryagin Differential Principle (PDP) mindset proposed in \cite{jin2020pontryagin}.



\section{Case Study: Cyber-physical Battery Systems}

Beyond the voltage stability scenario described in Section \ref{subsection: Pillar of D-STAB}, the fidelity-gap vulnerability, which forms the cornerstone of the D-STAB paradigm, also extends to other critical cyber-physical infrastructures, notably cyber-physical battery systems. Such systems necessitate diverse modeling complexities to precisely capture electrochemical behaviors within batteries. These complexities range from detailed Single Particle Models (SPM) to computationally efficient Equivalent Circuit Models (ECM).

In this section, we explore a representative scenario involving optimal battery charging governed by Battery Management Systems (BMS). This example serves to concretely illustrate how the D-STAB attack can leverage fidelity gaps to pose serious threats to battery infrastructures in practical operations.

BMS frequently utilizes ECMs for real-time decision-making due to their computational simplicity. However, ECMs inherently neglect detailed electrochemical dynamics, creating exploitable fidelity gaps. Here, we demonstrate how discrepancies between high-fidelity SPM and low-fidelity ECM enable an attacker to stealthily manipulate the charging process. Specifically, we employ a discretized SPM (using central differences and Euler's forward method) to design adversarial inputs.

\subsection{SPM-based High-fidelity Adversarial MPC}

The Single Particle Model (SPM) is a simplified yet widely used mathematical model for lithium-ion batteries \cite{perez2016optimal, park2017hybrid}. While it reduces the complexity of full electrochemical models, such as the Doyle-Fuller-Newman (DFN) model, it captures key dynamics of battery behavior, making it suitable for optimal control and real-time applications. In SPM, each electrode (anode and cathode) is represented by a single spherical particle, and the model focuses on the diffusion of lithium ions within these particles.

The SPM assumes that each electrode is represented by a single spherical particle, and the lithium concentration $c_s(r,t)$ varies radially according to Fick's second law of diffusion:
\begin{subequations}
    \label{eq: SPM diffusion}
    \begin{align}
        & \frac{\partial c_s^{\pm}(r,t)}{\partial t} = \frac{1}{r^2} \frac{\partial}{\partial r} \Big( D_s^{\pm} r^2 \frac{\partial c_s^{\pm}(r,t)}{\partial r}\Big), \label{eq: SPM diffusion 1}\\
        & \frac{\partial c_s^{\pm}(R_s,t)}{\partial r} = \mp \frac{1}{D_s^{\pm} F a^{\pm} L^{\pm}} I(t) \\
        & \frac{\partial c_s^{\pm}(0,t)}{\partial r} = 0
    \end{align}
\end{subequations} where $c_s^{\pm}(r,t)$ is the concentration in the solid particle at time $t$ (superscript ``$+$'' for the cathode, and ``$-$'' for the anode), $r \in [0, R_s]$ is the radial distance from the center to a point inside the particle, $R_s$ is the particle radius, and $D_s^{\pm}$ are the solid-phase diffusion coefficients. The terms $a^{\pm}$, $L^{\pm}$, and $I(t)$ represent the surface area, electrode thickness, and applied current, respectively.

In SPM, the State of Charge (SoC) is defined by both surface and bulk concentrations of lithium in the particles. The surface SoC is defined as the ratio of the lithium concentration at the particle surface to the maximum concentration, and the bulk SoC is the average concentration within the particle: $SoC_{\text{surf}}(t) = \frac{c_s^-(R_s, t)}{c_{s,\text{max}}^-}$, $SoC_{\text{bulk}}(t) = \frac{\bar{c}_s^-(t)}{c_{s,\text{max}}^-}$, where $\bar{c}_s^-(t) = \frac{3}{R_s^3} \int_{0}^{R_s} r^2c_s^-(r,t) \text{d} r$, and $c_{s,\text{max}}^{-}$ is the maximum lithium concentration in the anode.

Consider a discretized particle with 10 evenly spaced radial points indexed by $i \in [1, 10]$, where $i=1$ represents the center of the particle and $i=10$ is the surface of the particle. The spacing between these points is denoted as $\Delta r$, representing the radial step size. By using central difference approximation and Euler's forward method, we derive a high-fidelity adversarial MPC in the form of \eqref{eq: High fidelity adversarial MPC}, where the parameters $\theta_H:= \textbf{col}(q_H, r_H) \in \mathbb{R}^{2}$ are defined for the cost function, $(\textbf{c}_{s,1}^{-}(k|t),\cdots,\textbf{c}_{s,10}^{-}(k|t), \textbf{c}_{s,1}^{+}(k|t),\cdots,$
$\textbf{c}_{s,10}^{+}(k|t)):= x_{H, k|t} \in \mathbb{R}^{20}$ is the  high-fidelity system state at time step $t$, and prediction step $k$, $\Delta I_{k|t} = I_{k|t} - I_{k-1|t} := u_{k|t} \in \mathbb{R}$ is the incremental current as the control input, $g_H: \mathbb{R}^{20} \times \mathbb{R} \times \mathbb{R}^{2} \rightarrow \mathbb{R}$ is the instantaneous cost function, and $g_{H,N}: \mathbb{R}^{20} \times \mathbb{R}^{2} \rightarrow \mathbb{R}$ is the terminal cost function, which are defined as
\begin{subequations}
    \label{eq: high-order cost function}
    \begin{align}
        & \begin{split}
            g_H(x_{H, k|t}, u_{k|t}) = \Big\| \frac{3 \sum_{i=1}^{10} w_i r_i^2 x_{H,k|t}[i]}{R_s^3 c_{s,\text{max}}^-} - SoC_{\text d} \Big\|^2_{q_H} \\
            + \Big\| u_{k|t} \Big\|^2_{r_H}
        \end{split}\\
        & \begin{split}
            g_{H,N}(x_{H, N|t}) = \Big\| \frac{3 \sum_{i=1}^{10} w_i r_i^2 x_{H,N|t}[i]}{R_s^3 c_{s,\text{max}}^-} - SoC_{\text d} \Big\|^2_{q_H}
        \end{split},
    \end{align} 
\end{subequations} where $w_i$ is the weight for the $i$-th node, $w_1=\frac{1}{2}$, $w_2=1$, $\cdots$, $w_{9}=1$, $x_{H,k|t}[i]$ is the $i$-th element of vector $x_{H,k|t}$, $SoC_{\text d}$ is the desired SoC target, $w_{10}=\frac{1}{2}$, $r_i$ represents the radial position of the $i$-the node within the particle ($r_1 = 0$, and $r_N = R_s$).

$f_{H, k|t}: \mathbb{R}^{20} \times \mathbb{R} \rightarrow \mathbb{R}^{20}$ denotes the dynamics with initial condition $\frac{3 \sum_{i=1}^{10} w_i r_i^2 x_{H,0|0}[i]}{R_s^3 c_{s,\text{max}}^-} = SoC_{\text{bulk},\text{init}}$, which is defined as
\begin{equation}
    \label{eq: high-order dynamics}
    f_{H,k|t}(x_{H,k|t}, u_{k|t}) = A_H x_{H,k|t} + B_H u_{k|t} +B_H I_{k-1|t},
\end{equation} where 
$A_H = \begin{bmatrix}
        I + A_{\text{diff}}^+ & \textbf{0} \\
        \textbf{0} & I + A_{\text{diff}}^-
        \end{bmatrix}$, $B_H
        = \begin{bmatrix}
            B_{\text{diff}}^+ \\
            B_{\text{diff}}^-
          \end{bmatrix}$. \footnote{The detailed forms of $A_{\text{diff}}^{\pm}$ and $B_{\text{diff}}^{\pm}$ are omitted due to the space limit and can be found in the complete version on arXiv.}                     
A key aspect of the D-STAB attack is the design of high-fidelity adversarial constraints. In this case, we choose to design based on the concentration gradient constraint: $SoC_{\text{surf}}(k|t) - SoC_{\text{bulk}}(k|t) \leq \gamma_1 SoC_{\text{bulk}}(k|t) + \gamma_2$. 

\begin{remark}
    \label{remark: concentration gradient constraint}
    The constraint is pivotal as it regulates the concentration gradient between the surface and bulk regions to prevent excessive concentration gradients, which can result in lithium plating \cite{yang2017modeling} and SEI growth \cite{pinson2012theory}. While maintaining minimal gradients is ideal for battery longevity, the attack aims to maximize the difference, inducing harmful effects such as structural degradation.
\end{remark}

To maintain the stealthiness of the attack, additional high-fidelity constraints are imposed to mimic normal operation, including bounds on bulk SoC, current, and voltage, such as $SoC_{\text{bulk}}(k|t) \leq SoC_{\text{max}}$, $I(k|t) \leq I_{\text{max}}$, and $V(k|t) \leq V_{\text{max}}$.

\begin{remark}
    \label{remark: stealthiness of the attack}
    The stealthiness of the D-STAB attack is achieved through two mechanisms: 
    
    1) {\textbf{Mimicking Normal Operation}}: By replicating the low-fidelity states and constraints, the attacker ensures the system appears normal to the defender.;

    2) {\textbf{Exploiting the Fidelity Gap}}: The adversarial constraints leverage high-fidelity information that the defender cannot access, creating an information asymmetry that hides the attack.
\end{remark}

The corresponding $w_A: \mathbb{R}^{20} \rightarrow \mathbb{R}$, and $w_{H, k|t}: \mathbb{R}^{20} \times \mathbb{R} \rightarrow \mathbb{R}^3$ and $w_{H, N|t}: \mathbb{R}^{20} \rightarrow \mathbb{R}$ are defined as
\begin{equation}
    \label{eq: high-order adversarial inequality constraint}
    \begin{aligned}
        w_A(x_{H,k|t}) =& \frac{x_{H,k|t}[10]}{c_{s,\text{max}}^-} - (1 + \gamma_1) \frac{3 \sum_{i=1}^{10} w_i r_i^2 x_{H,k|t}[i]}{R_s^3 c_{s,\text{max}}^-} \\
        & - \gamma_2,  
    \end{aligned}
\end{equation}
and
\begin{subequations}
    \label{eq: high-order normal inequality constraints}
    \begin{align}
        & \begin{split}
            w_{H, k|t}(x_{N, k|t}, u_{k|t}) = \begin{bmatrix}
               \frac{3 \sum_{i=1}^{10} w_i r_i^2 x_{H,k|t}[i]}{R_s^3 c_{s,\text{max}}^-} - SoC_{\text{max}} \\
               u_{k|t} - I_{\text{max}} + I_{k-1|t} \\
               V(k|t) - V_{\text{max}}
            \end{bmatrix}
          \end{split}, \\
        & \begin{split}
            w_{N|t}(x_{N|t}) = \frac{3 \sum_{i=1}^{10} w_i r_i^2 x_{H,N|t}[i]}{R_s^3 c_{s,\text{max}}^-} - SoC_{\text{max}}
          \end{split},
    \end{align}
\end{subequations} respectively. The initial conditions for the high-fidelity MPC are given by $SoC_{\text{bulk}}(0|0) = SoC_{\text{bulk},\text{init}}, \ SoC_{\text{bulk}}(0|t) = SoC_{\text{bulk}}(t)$, and $I(-1) = I_{\text{init}}$.

\subsection{ECM-based Low-fidelity Target MPC}

The Rint model (or internal resistance model) is a simplified equivalent circuit model commonly used to represent the electrical behavior of rechargeable batteries, especially lithium-ion cells. In this model, the battery is described as an ideal voltage source (representing the open circuit voltage, OCV) in series with a resistor, which accounts for the internal resistance (Rint) of the cell. This internal resistance includes the ohmic resistance of the cell's materials and interfaces (e.g., electrode-electrolyte) as well as contributions from charge transfer and diffusion processes.

The battery's voltage response under load is modeled as the difference between the OCV and the voltage drop across the internal resistance, described mathematically as: 
\begin{equation}
    \label{eq: Rint state space representation}
    \begin{cases} 
        SoC_{k+1} = SoC_k + \frac{\eta \delta_k}{Q_c} I_k \\
        V_{k} = h_{OCV}(SoC_k) + R_0 I_k,
    \end{cases}
\end{equation} where $SoC_k$ denotes the state of charge at time step $k$, $\eta$ is the Coulombic efficiency, $\delta_k$ is the time interval between two consecutive time steps (in seconds), and $Q_c$ represents the battery capacity (in Ah). $I_k$ is the current at time step $k$ ($I_k > 0$ for charging and $I_k < 0$ for discharging). $V_k$ is the terminal voltage, $h_{OCV}(SoC_k)$ represents the OCV as a function of SoC, and $R_0$ is the internal resistance.

We can now formulate a low-fidelity MPC represented as \eqref{eq: Low-fidelity constrained OCP}, where the parameters $\theta:= {\textbf{col}(q, r)} \in \mathbb{R}^2$ represent the cost function weights, $SoC_{k|t} := x_{k|t} \in \mathbb{R}$ is the low-fidelity system state, and $\Delta I_{k|t} := u_{k|t} \in \mathbb{R}$ is the control input. The desired SoC target $SoC_{\text d}$ is denoted as $x^*$, and the instantaneous and terminal cost function $g: \mathbb{R} \times \mathbb{R} \times \mathbb{R}^2 \rightarrow \mathbb{R}$ and $g_N: \mathbb{R} \times \mathbb{R}^2 \rightarrow \mathbb{R}$ are defined as:
\begin{subequations}
    \label{eq: low-order MPC cost function}
    \begin{align}
        & g(x_{k|t}, u_{k|t};Q, R) = \|x_{k|t} - x^*\|^2_{Q} + \|u_{k|t}\|^2_{R}, \\
        & g_{N}(x_{N|t}; Q) = \|x_{N|t} - x^*\|^2_{Q}.
    \end{align}
\end{subequations} 

The system dynamics $f_{k|t}: \mathbb{R} \times \mathbb{R} \rightarrow \mathbb{R}$ with initial state $x_{0|t}$ are expressed as:
\begin{equation}
    \label{eq: low-order MPC system dynamics}
    f_{k|t}(x_{k|t}, u_{k|t}) = x_{k|t} + \frac{\eta \delta_t}{Q_c} u_{k|t} + \frac{\eta \delta_t}{Q_c} I_{k-1|t}.
\end{equation} 

The inequality constraints, $w_{k|t}: \mathbb{R} \times \mathbb{R} \rightarrow \mathbb{R}^3$ and terminal constraints $w_{N|t}: \mathbb{R} \rightarrow \mathbb{R}$  ensure the system remains within safe operating limits during the MPC operation. They are defined as:
\begin{subequations}
    \label{eq: low-order MPC inequality constraint}
    \begin{align}
        & \begin{split}
            & w_{k|t}(x_{k|t}, u_{k|t}) = \\
            & \begin{bmatrix}
               x_{k|t} - SoC_{\text{max}} \\
               u_{k|t} - I_{\text{max}} + I_{k-1|t} \\
               R_0 u_{k|t} - V_{\text{max}} + h_{OCV}(x_{k|t}) + R_0 I_{k-1|t}
            \end{bmatrix}
          \end{split} \\
        & \begin{split}
            w_{N|t}(x_{N|t}) = x_{N|t} - SoC_{\text{max}}
          \end{split}
    \end{align}
\end{subequations}

\subsection{Numerical Results and Analysis}

For the high-fidelity model, we reference the Single Particle Model (SPM) used in \cite{perez2016optimal}. Readers are encouraged to explore the corresponding GitHub repository \cite{moura2016spmet} for details on hyperparameter settings. To ensure consistency in the underlying physical processes, we identify key parameters for the Rint model (including battery capacity, OCV function, and internal resistance) based on data generated by the SPM model. This step is crucial for aligning the low-fidelity Rint model with the high-fidelity SPM, maintaining coherence between the models.

In the implementation of D-STAB, we set the adversarial constraint to deliberately violate the concentration gradient limit: $SoC_{\text surf}(k|t) - SoC_{\text bulk}(k|t) \geq \gamma_1 SoC_{\text bulk}(k|t) + \gamma_2$, where $SoC_{\text surf}(k|t)$ and $SoC_{\text bulk}(k|t)$ represent the normalized surface and average concentrations, respectively. This setting aligns with the adversarial target of D-STAB, which is to increase the concentration difference to accelerate battery degradation.

The design of the high-fidelity adversarial constraint depends on the selection of the hyperparameters $\gamma_1$ and $\gamma_2$. In this paper, we explore three levels of adversarial constraints: 1) Low-level, $\gamma_1 = 1e-2$, $\gamma_2 = 5e-4$; 2) Medium-level, $\gamma_1 = 4e-2$, $\gamma_2 = 1e-3$; 3) High-level, $\gamma_1 = 8e-2$, $\gamma_2 = 2e-3$. These levels represent the increasing difficulty of satisfying the designed adversarial constraints.

\begin{figure}[htbp]
\centering
\includegraphics[width=0.48\textwidth]{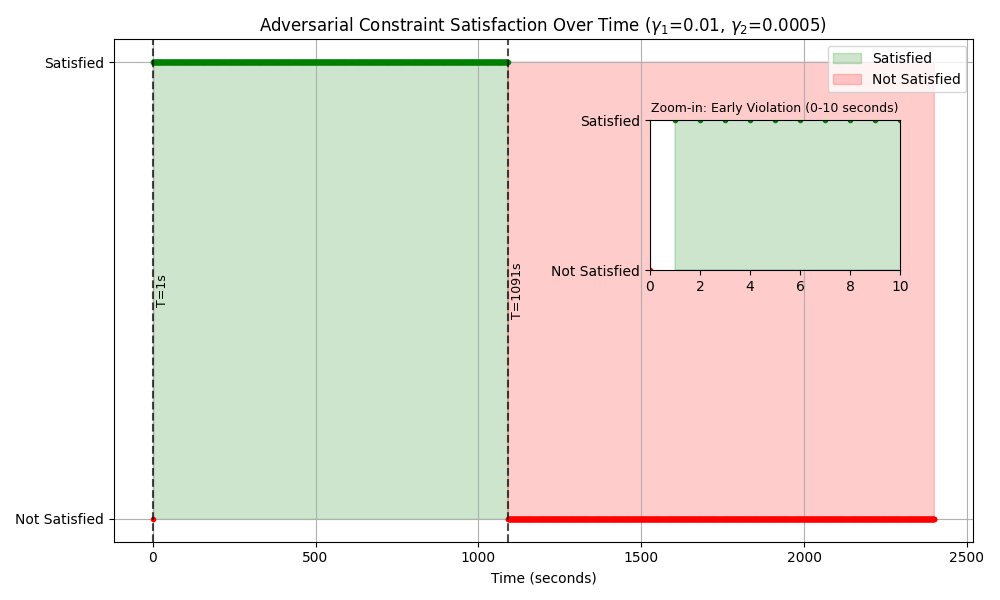}
\caption{Low-level Adversarial Constraint}
\label{fig: Low-level Adversarial Constraint}
\end{figure}

\begin{figure}[htbp]
\centering
\includegraphics[width=0.48\textwidth]{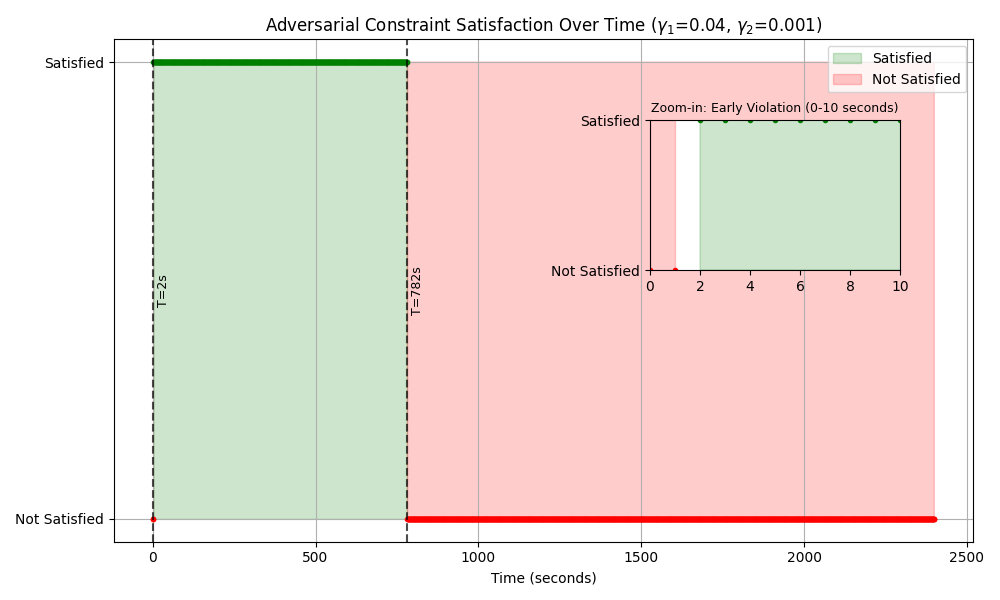}
\caption{Medium-level Adversarial Constraint}
\label{fig: Medium-level Adversarial Constraint}
\end{figure}

\begin{figure}[htbp]
\centering
\includegraphics[width=0.48\textwidth]{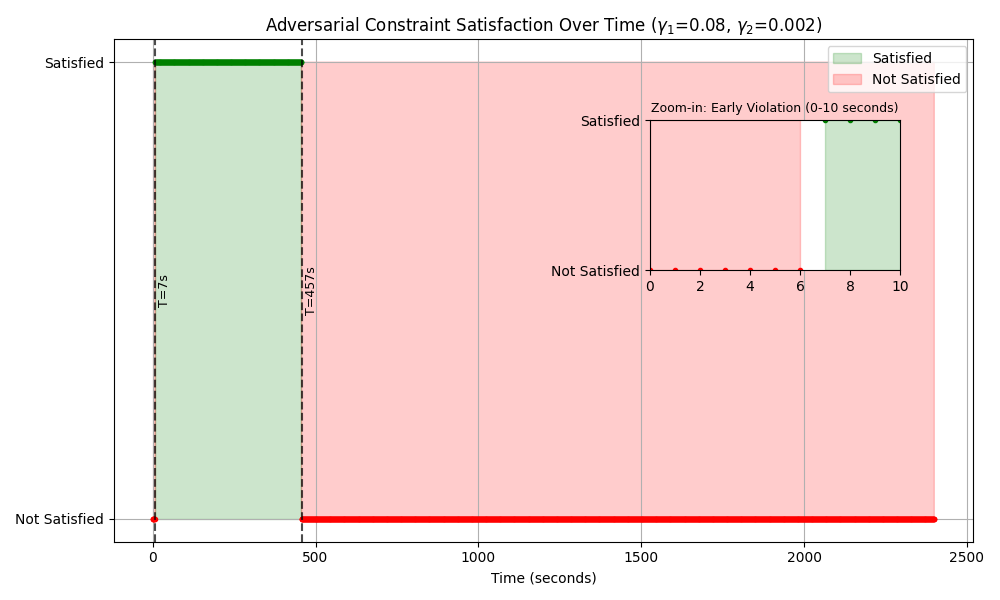}
\caption{High-level Adversarial Constraint}
\label{fig: High-level Adversarial Constraint}
\end{figure}

After deriving the ``optimal'' adversarial weighting parameters, $q^*$ and $r^*$, we substitute them into the Rint model-based low-fidelity MPC and solve for the output control sequence. By testing the derived control sequence in the corresponding high-fidelity adversarial constraints, the results are as shown in Fig.~\ref{fig: Low-level Adversarial Constraint}, Fig.~\ref{fig: Medium-level Adversarial Constraint} and Fig.~\ref{fig: High-level Adversarial Constraint}, respectively.

From these results, we observe the effectiveness of the attack and that as the adversarial constraints become stricter, the number of time steps where the constraint is satisfied decreases. This highlights a fundamental tradeoff between the strength of the attack and its active duration. Finding the optimal balance is essential for maximizing the effectiveness of the attack in real-world applications.

\section{Concluding Remarks}

In this paper, we proposed the Dual State-Space Fidelity Blade (D-STAB) paradigm, a stealthy cyber-physical attack strategy exploiting fidelity gaps between high- and low-fidelity models in cyber-physical systems. By precisely manipulating firmware parameters using high-fidelity state-space information, D-STAB can stealthily induce severe operational disruptions. Through our case study on optimal battery charging within Battery Management Systems, we demonstrated D-STAB’s capability to stealthily manipulate critical infrastructure operations. Future research should prioritize the development of advanced detection and mitigation techniques aimed at reducing fidelity gaps.





\bibliographystyle{IEEEtran}
\bibliography{main}

\end{document}